%% file: ops10_arXiv.tex
\documentclass[
	twocolumn,
	aps,
	prd,
	superscriptaddress,
	preprintnumbers,
	secnumarabic,
	nofootinbib]{revtex4-1}

\pdfoutput=1

\usepackage{amsmath}         
\usepackage{amssymb}         
\usepackage{amsfonts}        
\usepackage{graphicx}        

\usepackage{mathrsfs} 
\usepackage{verbatim}
\usepackage{float}
\usepackage{slashed}
\usepackage{bbm}
\usepackage[english]{babel}
\usepackage{color}
\usepackage{xcolor}
\usepackage{xspace}
\usepackage{enumerate}
\usepackage{url}
\usepackage{bbding}

\newcommand{\acro}[1]{\textsc{\MakeLowercase{#1}}}    
\newcommand{\DM}{\acro{DM}\xspace}
\newcommand{\JWST}{\acro{JWST}\xspace}

\renewcommand{\tilde}{\widetilde}   
\renewcommand{\vec}[1]{\mathbf{#1}} 
\newcommand{\Xmark}{\text{\sffamily X}}        

\input{universalnewcommands.tex}

\def\mdm{m_\chi}
\def\rhodm{\rho_\chi}
\def\vhalo{v_{\rm halo}}
\def\sigmanDM{\sigma_{\chi n}}
\def\sigmathresh{\sigma_{t}}
\def\sigmageom{\sigma_0}
\def\cutoff{\Lambda}

\def\Er{E_R}
\def\KEhalo{{\rm KE}_{\rm halo}}

\def\Mss{\Oc_{\tt SS}}
\def\Mpp{\Oc_{\tt PP}}
\def\Msp{\Oc_{\tt SP}}
\def\Maa{\Oc_{\tt AA}}

\usepackage{outlines}

\definecolor{darklightsabergreen}{rgb}{0.0, .49, 0.06}
\definecolor{orange}{rgb}{1.0, 0.5, 0.0}

\begin{document}
\preprint{\texttt{UCR-TR-2017-FLIP-NCC-1701}}

\title{Neutron stars at the dark matter direct detection frontier}

\author{Nirmal Raj}
\email{\texttt{nraj@nd.edu}}
\affiliation{Department of Physics, University of Notre Dame, 225 Nieuwland Hall, Notre Dame, IN 46556, USA}

\author{Philip Tanedo}
\email{\texttt{flip.tanedo@ucr.edu}}
\affiliation{Department of Physics \& Astronomy, \\ University of California, Riverside, CA 92521, USA}

\author{Hai-Bo Yu}
\email{\texttt{haiboyu@ucr.edu}}
\affiliation{Department of Physics \& Astronomy, \\ University of California, Riverside, CA 92521, USA}

\begin{abstract}
Neutron stars capture dark matter efficiently.
The kinetic energy transferred during capture heats old neutron stars in the galactic halo to temperatures detectable by upcoming infrared telescopes.
We derive the sensitivity of this probe in the framework of effective operators.
For dark matter heavier than a GeV, we find that neutron star heating can set limits on the effective operator cutoff that are orders of magnitude stronger than possible from terrestrial direct detection experiments in the case of spin-dependent and velocity-suppressed scattering.
\end{abstract}

\maketitle


\section{Introduction}

Astrophysical and cosmological data imply the existence of dark matter (\DM), but its particle properties remain hidden from terrestrial experiments. 
Contact operators are a useful parameterization of the underlying dynamics when the transfer momentum $\vec{q}$ is small, such as in direct detection experiments where \DM scatters off target nuclei.
The contact operators highlight the sensitivity of direct detection to the structure of the interaction between the dark and visible sectors.

For example, \emph{nuclear coherence} enhances a spin-independent cross-section through a vector--vector operator by seven orders of magnitude compared to a spin-dependent axial vector--axial vector operator.
Moreover, the spin-dependent pseudoscalar--pseudoscalar operator is suppressed by four powers of the \emph{small momentum transfer}, $\vec{q}^4/m_\chi^2m_n^2$, relative to the axial--axial operator~\cite{0908.3192,1008.1591,1012.5317,1203.3542,1305.0912,1305.1611,1312.7772}.
There is thus a hierarchy of sensitivity in the types of \DM dynamics encoded by effective operators that describe \DM scattering with nuclear targets.

\begin{figure}
 \includegraphics[width=.45\textwidth]{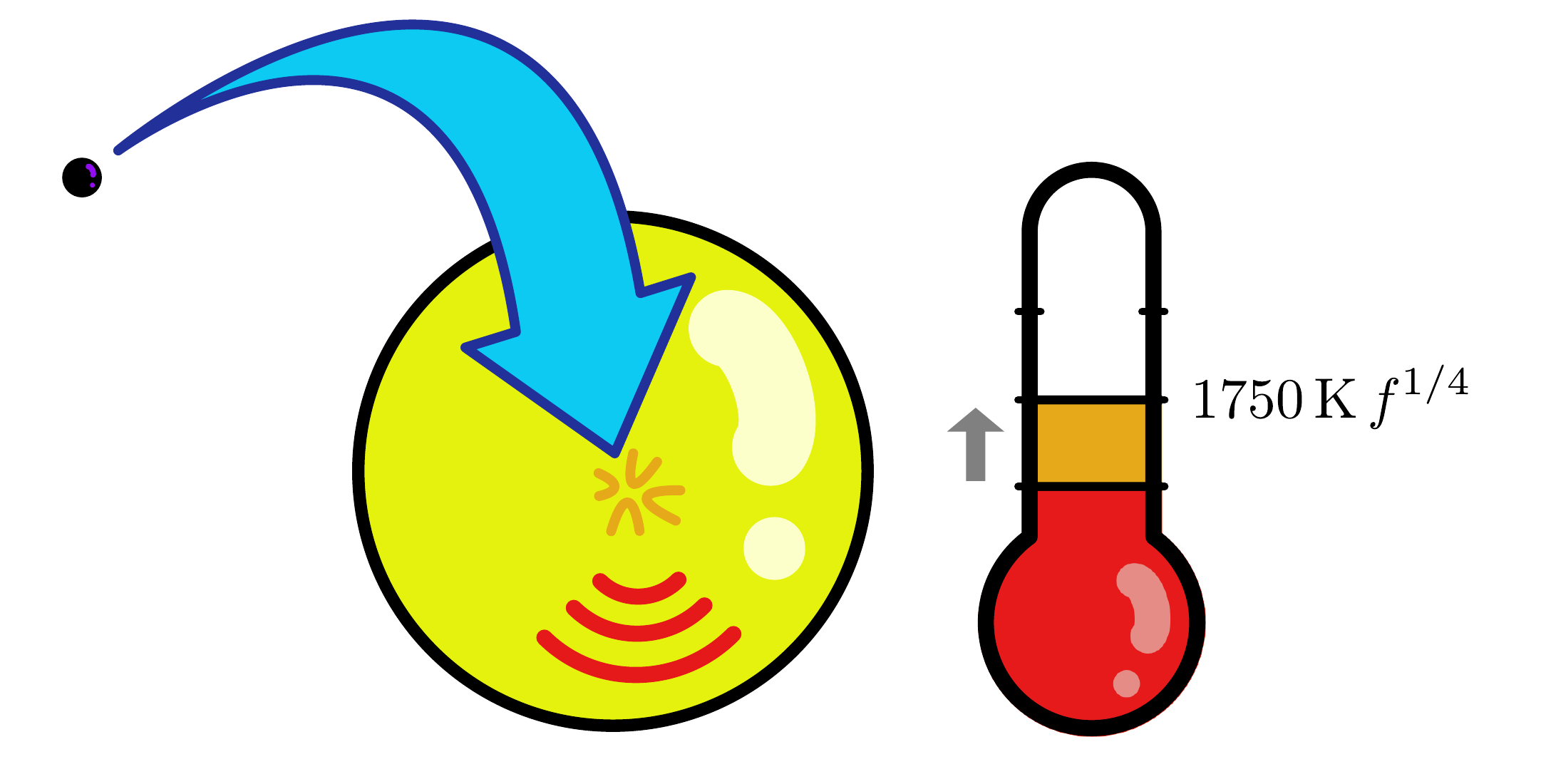} 
 \caption{\label{fig:schematic} 
Ambient dark matter is accelerated as it approaches a neutron star. This flux of kinetic energy is deposited into the star as heat, predicting a warmer neutron star temperature than in the absence of a dark matter--Standard Model interaction.
 }
 \end{figure}

Neutron stars are efficient targets for dark matter. 
The dark matter capture rate is largely agnostic to whether an interaction is spin-dependent or spin-independent, and the gravitational acceleration to $\mathcal O(0.5 c)$ speeds washes out velocity-suppression.
They were previously examined as laboratories to study \DM self-interactions and primordial asymmetries by considering capture followed by either annihilation or stellar implosion~\cite{Goldman:1989nd,0708.2362,1004.0586,1004.0629,1012.2039,1103.5472,1201.2400,1301.0036,1301.6811,1310.3509,1405.1031,1504.04019,1703.04043,1706.00001}.
More recently, \cite{1704.01577} demonstrated that \DM \ {scattering} alone may kinetically heat neutron stars to infrared temperatures that are detectable by next-generation infrared telescopes such as the James Webb Space Telescope (\JWST). This process is depicted in Fig.~\ref{fig:schematic}.
Measuring the temperature of even a single old, isolated neutron star $\sim$10 parsecs from the Sun is sufficient to obtain bounds on the \DM scattering cross-section.

Unlike conventional \DM probes such as (in)direct detection and colliders, neutron star heating is less dependent on \DM properties: low masses are not limited by the recoil threshold of direct detection, nor is there a mass cutoff above which the missing energy searches at colliders are ineffective.
Kinetic heating is independent of whether the dark matter ultimately annihilates, so that it is even robust against asymmetric \DM scenarios where there are no indirect detection signals.

In this note, we map the reach of a neutron star temperature observation on the coefficients of an effective contact operator basis that connects dark matter and quark bilinears. We select an illustrative set of operators that cover the range of spin- and momentum-dependence in scattering. We show that neutron star heating improves the reach on this parameter space by orders of magnitude compared to terrestrial direct detection.

\section{Dark Heating of Neutron Stars}
\label{sec:heating}

\paragraph*{Signals from neutron stars.}
 A typical neutron star has mass and radius
\begin{align}
	M_\star &= 1.5~M_\odot \ 
	&&\text{and}& 
	R &= 10~\text{km} \ .
	\label{eq:ns:mass:and:radius}
\end{align} 
It accelerates dark matter of mass $\mdm$ to kinetic energies of $(\gamma - 1) \mdm \sim 0.35 \mdm$. 
Assuming \DM densities and velocities typical of the solar system, $\rhodm = 0.4 ~{\rm GeV/cm}^3$ and $\vhalo = 230~{\rm km}/{\rm s}$, the flux of dark matter kinetic energy onto the neutron star is $25~\text{g}/\text{sec}$.
This maintains the neutron star at an infrared equilibrium blackbody temperature.
By comparison, in the absence of this mechanism, a neutron star older than $10^8$ years is expected to have cooled to a temperature that is $\Oc(10)$ lower \cite{Yakovlev:2004iq}, meaning the ``dark kinetic heating'' signal from such stars is essentially free of internal backgrounds.

\paragraph*{Energy deposition.}
The dark matter contribution to the neutron star temperature is \cite{1704.01577}, 
\begin{align}
T_{\rm kin} &= 1750~f^{1/4}~\!\text{K}
\ ,
\label{eq:temperatures}
\end{align}
where $f$ is the \DM capture efficiency. It depends on the ratio of the \DM--nucleon cross section $\sigmanDM$ to a threshold cross section $\sigmathresh$ above which all transient \DM captures,
\begin{align}
f &= {\rm min}(\sigmanDM/\sigmathresh,1)~.
\label{eq:efficiency}
\end{align}
The threshold cross section, in turn, is proportional to the neutron star geometric cross section $\sigmageom = \pi (m_n/M_\star) R^2 \simeq 2 \times 10^{-45} \ {\rm cm}^2$,
\begin{align}
\displaystyle
\sigmathresh =
 \begin{cases}
	\frac{\text{GeV}}{\mdm} \,\sigmageom
	&\text{if}~~\mdm < \text{GeV}, 
	\\
	\sigmageom
	&\text{if}~~\text{GeV}\leq\mdm \leq 10^6~\!\text{GeV}, 
	\\
	\frac{\mdm}{10^6~{\rm GeV}}\,\sigmageom
	&\text{if}~~\mdm > 10^6~{\rm GeV}.
 \end{cases}
 \label{eq:sigmathreshold}
\end{align}
The mass-dependence of this relation comes from the typical recoil energy which, in the neutron rest frame, is
\begin{align}
\Er &= \frac{m_n \mdm^2 \gamma^2 v^2_{\rm esc}}{(m_n^2+\mdm^2+2\gamma m_n\mdm)}~.
\label{eq:Erec}
\end{align}
\begin{enumerate}
	\item For $\mdm \!<\!\text{GeV}$, 
	the typical momentum transfer $\sqrt{2m_n \Er}$ is smaller than the neutron star Fermi momentum $p_F \simeq 0.45~\text{GeV} \left[\rho_{\rm NS}/(4 \times 10^{38}~\text{GeV}\,\text{cm}^{-3})\right]$.
	Pauli blocking from the degenerate neutrons restricts scattering to the fraction, $3 \sqrt{2m_n \Er}/ p_F$,
		of neutrons that are close enough to the Fermi surface so that
 	$\sigmathresh \propto \Er^{-1/2} \propto \mdm^{-1}$.

	\item For $\text{GeV} \!\leq\! \mdm \!\leq\! 10^6~\!\text{GeV}$, a single scattering with $\Er \simeq m_n v_{\rm esc}^2 \gamma^2$ depletes \DM of its halo kinetic energy, $\mdm \vhalo^2/2$, and gravitationally binds it to the neutron star. Thus $\sigmathresh = \sigmageom$.

	\item For $\mdm > 10^6\!~\text{GeV}$, the \DM halo kinetic energy exceeds the recoil energy imparted to the neutron so that capture requires multiple scatters. The threshold cross section is proportional to the number of scatters, 
		$\sigmathresh\propto \KEhalo/\Er \propto \mdm$. 
\end{enumerate}

The observation of one or more old neutron stars at a given temperature determines $f$ through Eq.~\ref{eq:temperatures}. 
Given $f$, one may infer the \DM--nucleon cross section as a function of the \DM mass through Eqs.~\ref{eq:efficiency} and \ref{eq:sigmathreshold}, which can, in turn, be recast to model parameters.
\begin{table*}[hbt]
\begin{center}
\begin{tabular}{l l l l l c}
\hline 
\hline
Name  
& Operator 
& Coupling
& Matrix element
& $c(\vec{q})$ 
& Nuclear coherence
\\ 
\hline
$\Mss$ 
& $\displaystyle (\bar{\chi}\chi)(\bar{q}q)$ 
& $\displaystyle {y_q}/{\cutoff^2}$ 
& $4 \mdm m_n $ 
& $\displaystyle \frac{1}{\pi}$ 
& \Checkmark
\\
$\Maa$ 
& $\displaystyle (\bar{\chi}\gamma_5\gamma^\mu \chi)(\bar{q}\gamma_5\gamma_\mu q)$  
& $\displaystyle {1}/{\cutoff^2}$  
& $16\mdm m_n(\vec{S_\chi}\cdot\vec{S_n})$ 
& $\displaystyle \frac{3}{4\pi}$  
& \Xmark
\\
$\Msp$ 
& $\displaystyle  (\bar{\chi}\chi)(\bar{q}\gamma_5q)$ 
& $\displaystyle  {y_q}/{\cutoff^2}$ 
& $4\mdm (\vec{q}\cdot\vec{S_n})$ 
& $\displaystyle\frac{1}{16\pi}\frac{\mathbf{q}^2}{m^2_n}$  
& \Xmark
\\
$\Mpp$ 
& $\displaystyle (\bar{\chi}\gamma_5\chi)(\bar{q}\gamma_5 q)$ 
& $\displaystyle  {y_q}/{\cutoff^2}$ 
& $- 4 (\vec{q}\cdot\vec{S_\chi})(\vec{q}\cdot\vec{S_n})$ 
& $\displaystyle \frac{1}{64\pi} \frac{\mathbf{q}^4}{\mdm^2 m^2_n}$  
& \Xmark
\\
\hline
\end{tabular}
\end{center}
\caption{Operators considered in this work. The third column is the effective coupling, with $\Lambda$ the cutoff scale on which bounds are set and $y_q$ the quark Yukawa coupling as required by minimal flavor violation.
The fourth column is the scattering matrix element that encapsulates the spin and momentum dependence of each operator.
The fifth column provides the pre-factor in the \DM-nucleon cross section in Eq.~\ref{eq:DDXS}.
The sixth column states whether an operator permits coherent scattering across nuclei in direct detection experiments.
}
\label{tab:operators}
\end{table*}


\section{Contact Operators}

\DM scattering with ordinary matter can be parameterized by a set of contact operators when the transfer momentum is small compared to any intermediate particles.
We assume that \DM is a Majorana fermion that interacts with quarks through a basis of dimension-6 operators. 
The Lorentz structure of the fermion bilinears determines the spin and velocity dependence of the scattering matrix element~\cite{1305.1611}.
For example, to leading order in the \DM velocity, 
$\langle n|\bar{q} q| n\rangle=2m_n$, which is both spin- and velocity-{\em independent}, whereas $\langle n|\bar{q} \gamma_5 q|n\rangle=2(\vec{q}.\vec{S_n})$,  which is spin- and velocity-{\em dependent}. 

We apply the analysis of Sec.~\ref{sec:heating} to the four contact operators in Table~\ref{tab:operators}.
These operators span the behavior of spin and momentum dependence in \DM--nucleus scattering.
We assume minimal flavor violation \cite{Chivukula:1987py,Hall:1990ac,Buras:2000dm,MFV1} to ensure compatibility with stringent bounds on flavor-violating observables. Thus spin-0 Standard Model bilinears are proportional to quark masses and spin-1 bilinears are flavor-universal.
Up to this flavor proportionality, the couplings of the contact operators are encoded in the cutoff scale, $\Lambda$.

For a given operator, the \DM--neutron scattering cross section is
\begin{align}
\sigmanDM &= \ c(\vec{q})~\mu^2_{\chi n}~|f_n^\mathcal{O}|^2~,
\label{eq:DDXS}	
\end{align}
where 
$\mu_{\chi n}$ is the \DM--neutron reduced mass, and $c(\vec{q})$ encapsulates the transfer momentum dependence, and $f_n$ is the \DM--neutron coupling. The $c(\vec{q})$ are listed in Table~\ref{tab:operators}.
The $f_n$ are  
\begin{align}
f_n^{\tt SS} &= \nn \frac{\sqrt{2}m_n}{v \Lambda^2} \left( \sum_{q = u, d, s} f^{(n)}_{T_q} + \sum_{Q=c,b,t} \frac{2}{27} f^{(n)}_{T_G}\right)
\\
f_n^{\tt AA} &= 
\nn \frac{1}{\Lambda^2}  \left( \sum_{q = u, d, s} \Delta^{(n)}_q \right)
\\
f_n^{\tt SP, PP} &= \nn \frac{\sqrt{2}}{v\Lambda^2} \left( \sum_{q=u,d,s} m_q \Delta^{(n)}_{q_0} - \sum_{Q = c, b, t} \tilde{G}^{(n)}_0  \right)~,
\end{align}
where $v$ = 246~GeV is the electroweak vev, and the  matrix element coefficients on the right-hand side, estimated in lattice \acro{QCD}, are tabulated in many sources, e.g.~\cite{1012.5317,1312.7772}.
Analogous expressions hold for \DM--proton scattering in direct detection.
In Eq.~\ref{eq:DDXS} we fix ${\bf q}$ to a typical reference momentum transfer
 ${\bf q_{\rm ref}}$ 
~\cite{1012.5317}.
For direct detection experiments, ${\bf q}^2_{\rm ref} = 2 m_T E_{R_T}$, where $m_T$ is the mass of the target nucleus and $\Er^T = \mu^2_{T\chi}\vhalo^2/m_T$ is the typical nuclear recoil energy with $\mu_{T\chi}$ the nuclear target--\DM reduced mass.
This is in contrast to the recoil energy for scattering in a neutron star, Eq.~\ref{eq:Erec}. 

In contrast to terrestrial searches, neutron star heating constrains the operators $\Mss, \Maa$ and $\Msp$ with sensitivities comparable to each other.
Because \DM scatters directly with neutrons, the threshold cross-section in Eq.~\ref{eq:sigmathreshold} applies to both spin-independent and spin-dependent scattering. Momentum transfers are typically comparable to the nucleon mass, and there is no hierarchy based on velocity dependence.
For the operator $\Mpp$, the sensitivity for $\mdm$ above the neutron mass must fall as ${\bf q}^2/\mdm^2$ due to the $\bar\chi\gamma^5\chi$ bilinear.

\section{Annihilation \& Time Scales}
\label{sec:ann}

 \begin{figure}
 \includegraphics[width=.44\textwidth]{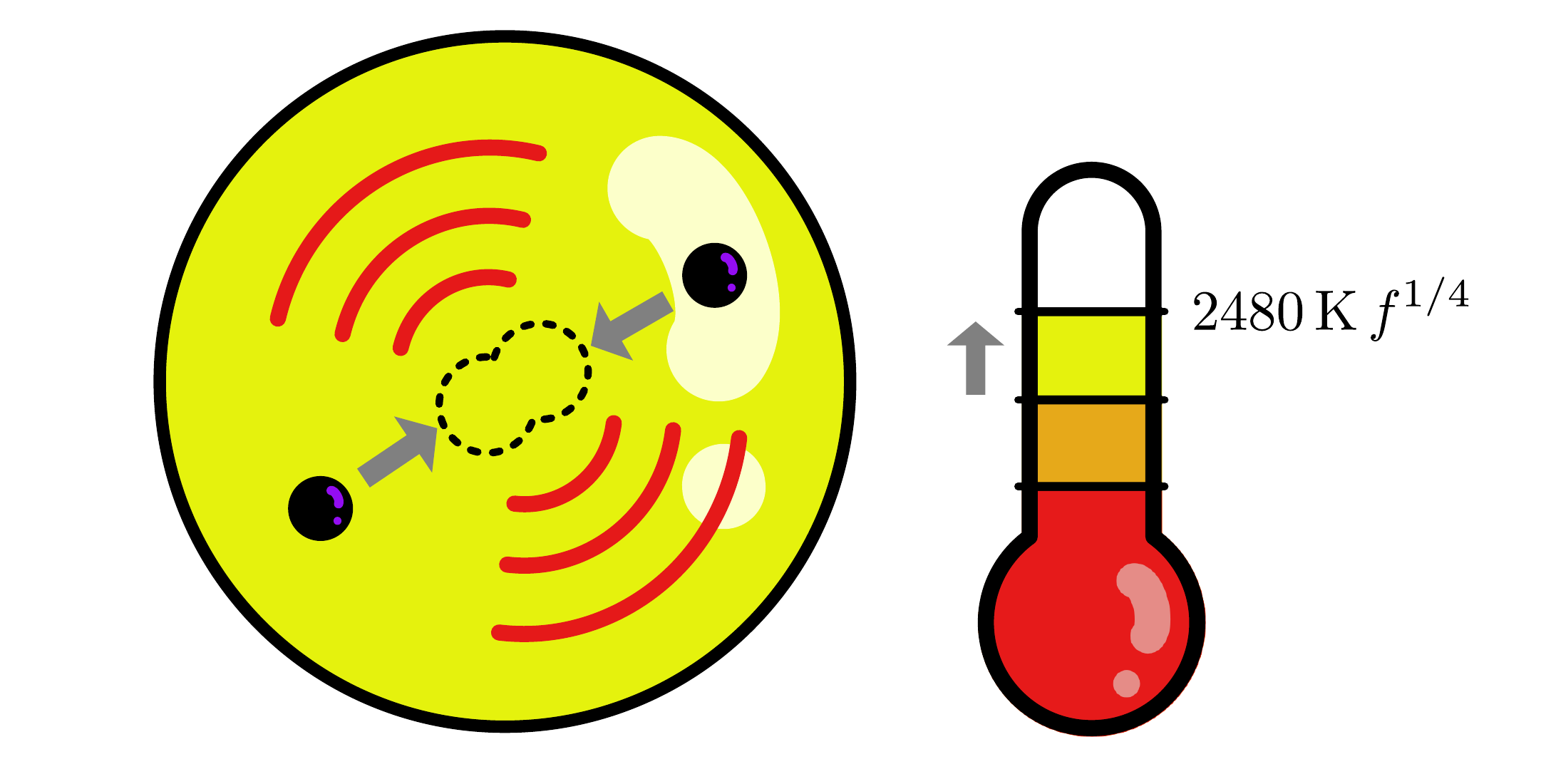}
 \caption{\label{fig:ann} Captured dark matter may annihilate and heat a neutron star. This contributes an additional energy flux that raises the neutron star temperature.
 }
 \end{figure}

Crossing symmetry relates the elastic scattering cross section to the \DM annihilation rate inside the neutron star.
The energy released in the process can raise the stellar temperature, Fig.~\ref{fig:ann},
\begin{align}
T_{\rm ann} &= 2480~f^{1/4}~\!\text{K}
\ .
\label{eq:ann:temperature}
\end{align}
In order to realize this temperature, the age of the neutron star $t_\text{NS}$ must exceed the combined time for captured dark matter to thermalize with the stellar core $t_\text{therm}$ and to equilibrate with the capture rate $t_\text{eq}$,
\begin{align}
	t_\text{NS} > t_\text{therm} + t_\text{eq} \, .
	\label{eq:annihilation:allowed}
\end{align}
The operators $\Mss$ and $\Maa$ satisfy $t_{\rm therm} \lsim 10^8$~years and $t_{\rm eq} < 10^6$~years for $f \geq 0.025$~\cite{1004.0586,1309.1721}. Since the neutron stars relevant for this study are older than $10^8$~years, these operators produce a stellar temperature of $T_{\rm ann}$ as opposed to $T_{\rm kin}$ in Eq.~\ref{eq:temperatures}. 
In contrast, the velocity-dependent operators $\Msp$ and $\Mpp$, require a more extensive study beyond the treatment in \cite{1309.1721} and we leave it for future work. 

The warmer temperature from \DM annihilation in addition to kinetic heating shortens the required telescope exposure time $t_{\rm obs}$ by a factor of 10~\cite{1704.01577}.
Observing a $T_{\rm ann} = 2480$~K neutron star using the F200W filter of the \acro{NIRC}am imager at a signal-to-noise ratio (\acro{SNR}) of 2 requires an exposure of $t_{\rm obs} = 9000~{\rm s}~(d/10{\rm pc})^4$, where $d$ is the distance from Earth.  
Using the \JWST pocket guide \cite{JWSTPG}, we find that for observing a $T_{\rm ann} = 1000$~K (peak wavelength = 2.9 $\mu$m) neutron star, the optimal filter is F356W (centered at 3.6 $\mu$m), which gathers 2.2~nJy at 2~\acro{SNR} in $10^4~$s. 
This translates to $t_{\rm obs} = 6 \times 10^6~{\rm s}~(d/10{\rm pc})^4$.
Observation times of this scale are obtainable in a potential deep field survey.

\section{Results}
\label{sec:results}

Neutron star heating sets upper limits on the cutoff scale of the effective operators, $\Lambda$.
We assume that \DM--quark interactions are dominated by a single contact operator and take the limit where the neutron is a point particle.
The point-like neutron limit reflects the assumption that neutron matrix elements of quark currents are proportional to that of the corresponding neutron currents $\langle n|\bar{q} \Gamma_q q|n \rangle = \Delta q^{(n)} \langle n| \bar{n} \Gamma_q n |n \rangle$, where $\Delta q^{(n)}$ is determined from experiment or the lattice~\cite{1312.7772}. 
It is also a practical limit given the unreliability of the neutron's parton distribution functions at $Q^2 \leq (1~{\rm GeV})^2$ \cite{Jaffe:1989jz,1405.7370}.
We quote limits assuming the benchmark neutron star with mass and radius in Eq.~\ref{eq:ns:mass:and:radius};
varying these properties affects stellar luminosities, and hence apparent temperatures and telescopic detection times~\cite{1704.01577}. 

\begin{figure*}
\begin{center}
\includegraphics[width=.47\textwidth]{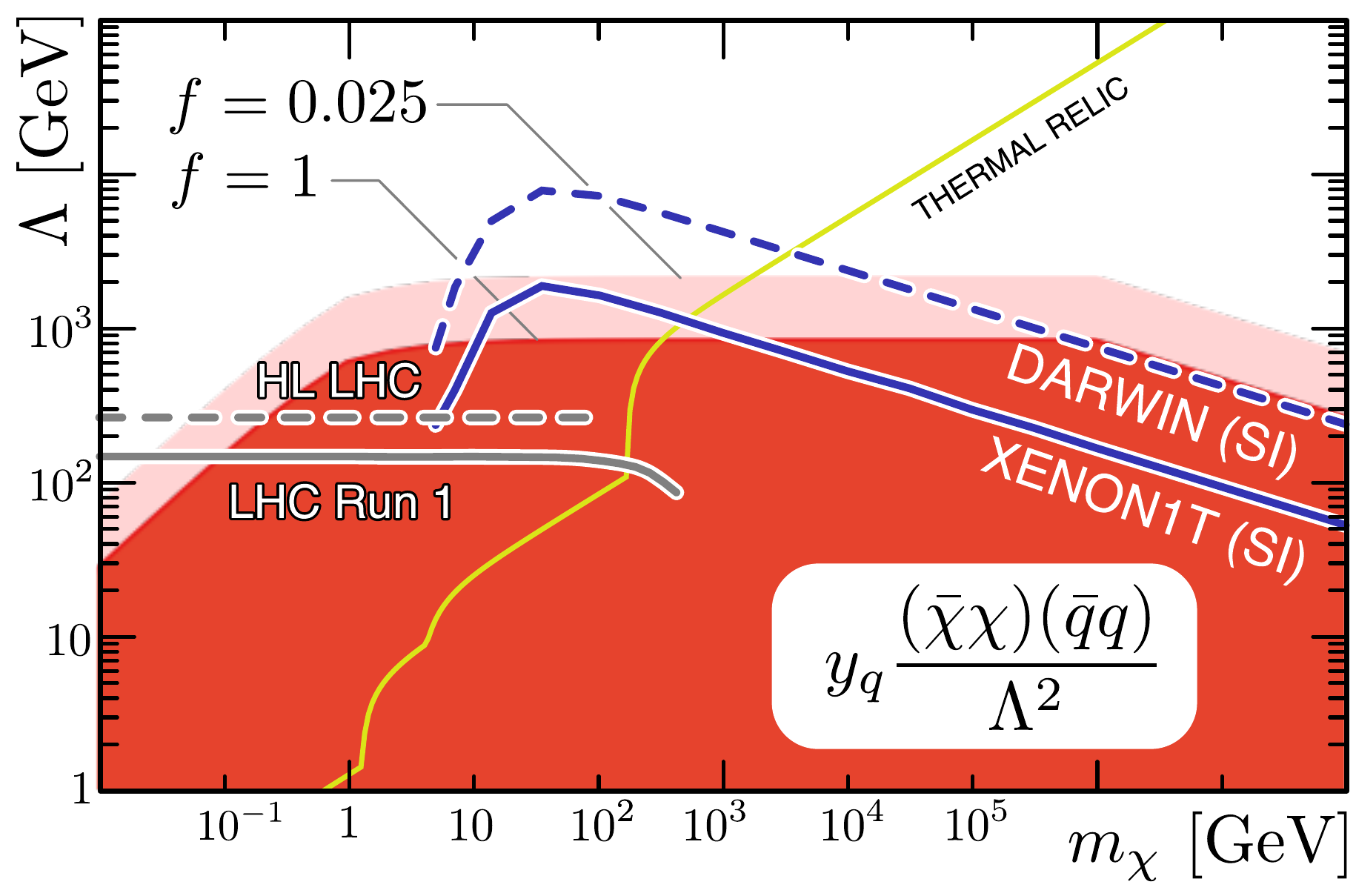} \qquad
\includegraphics[width=.47\textwidth]{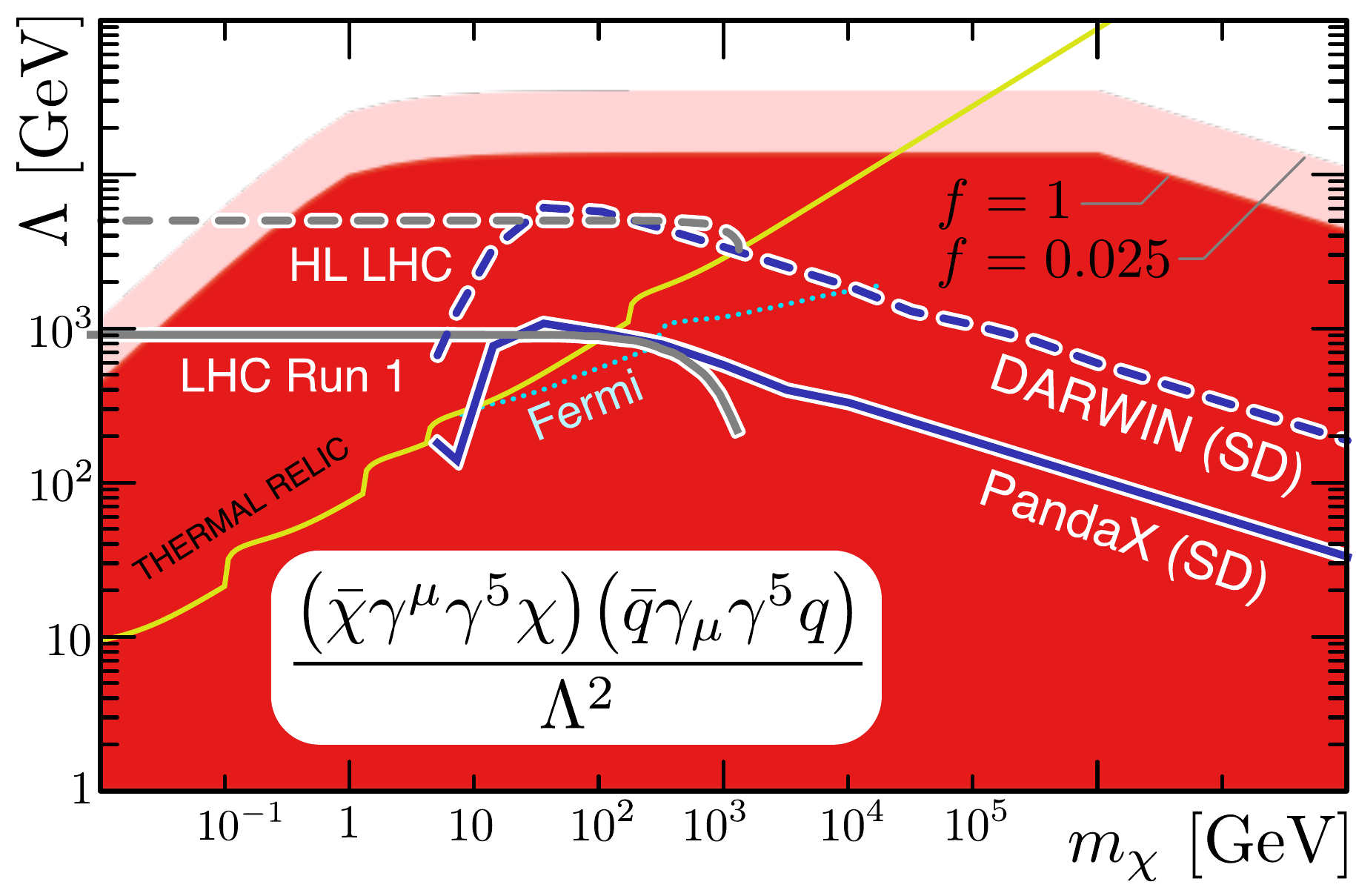} 
\\ 
\includegraphics[width=.47\textwidth]{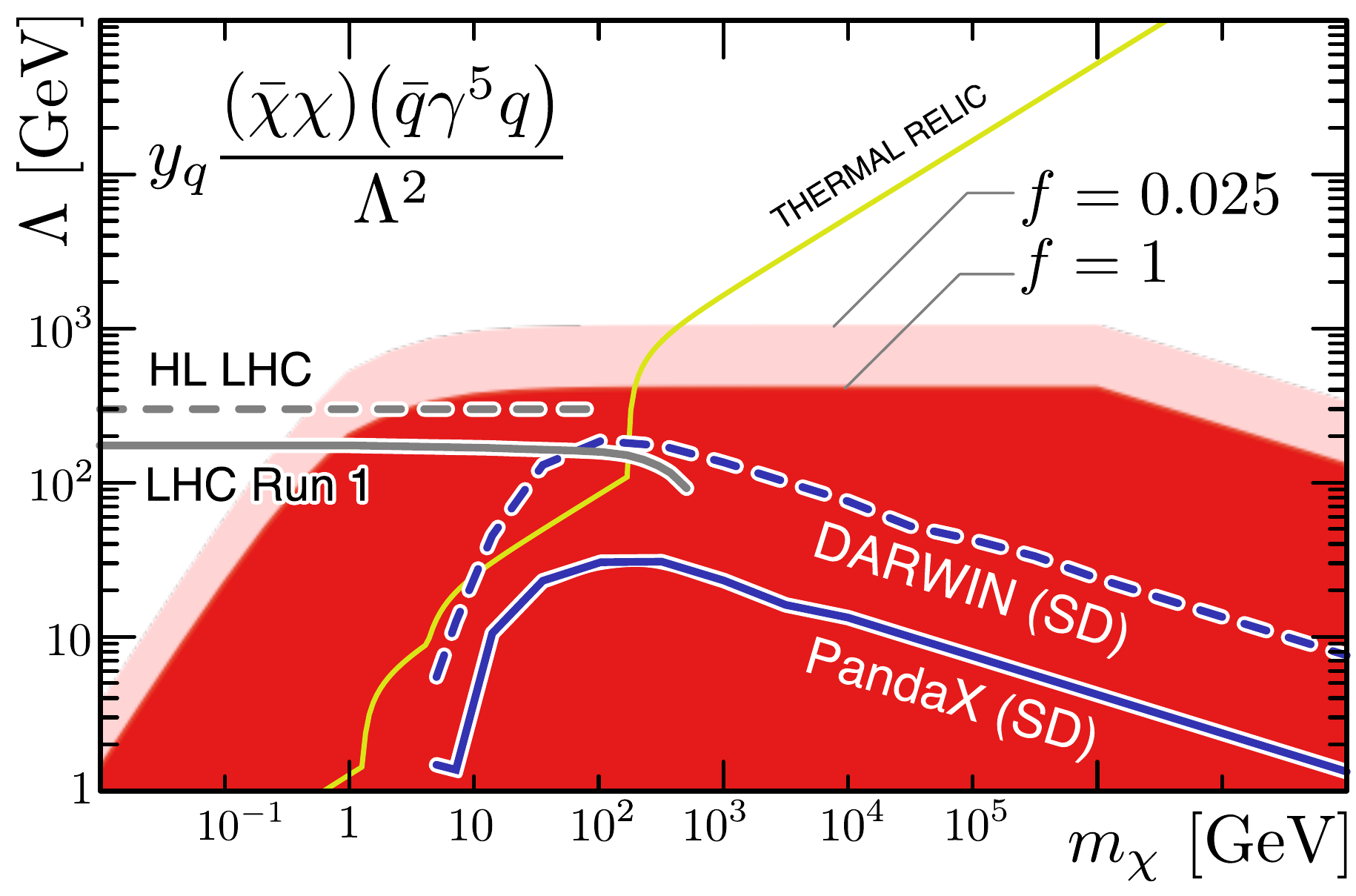} \qquad
\includegraphics[width=.47\textwidth]{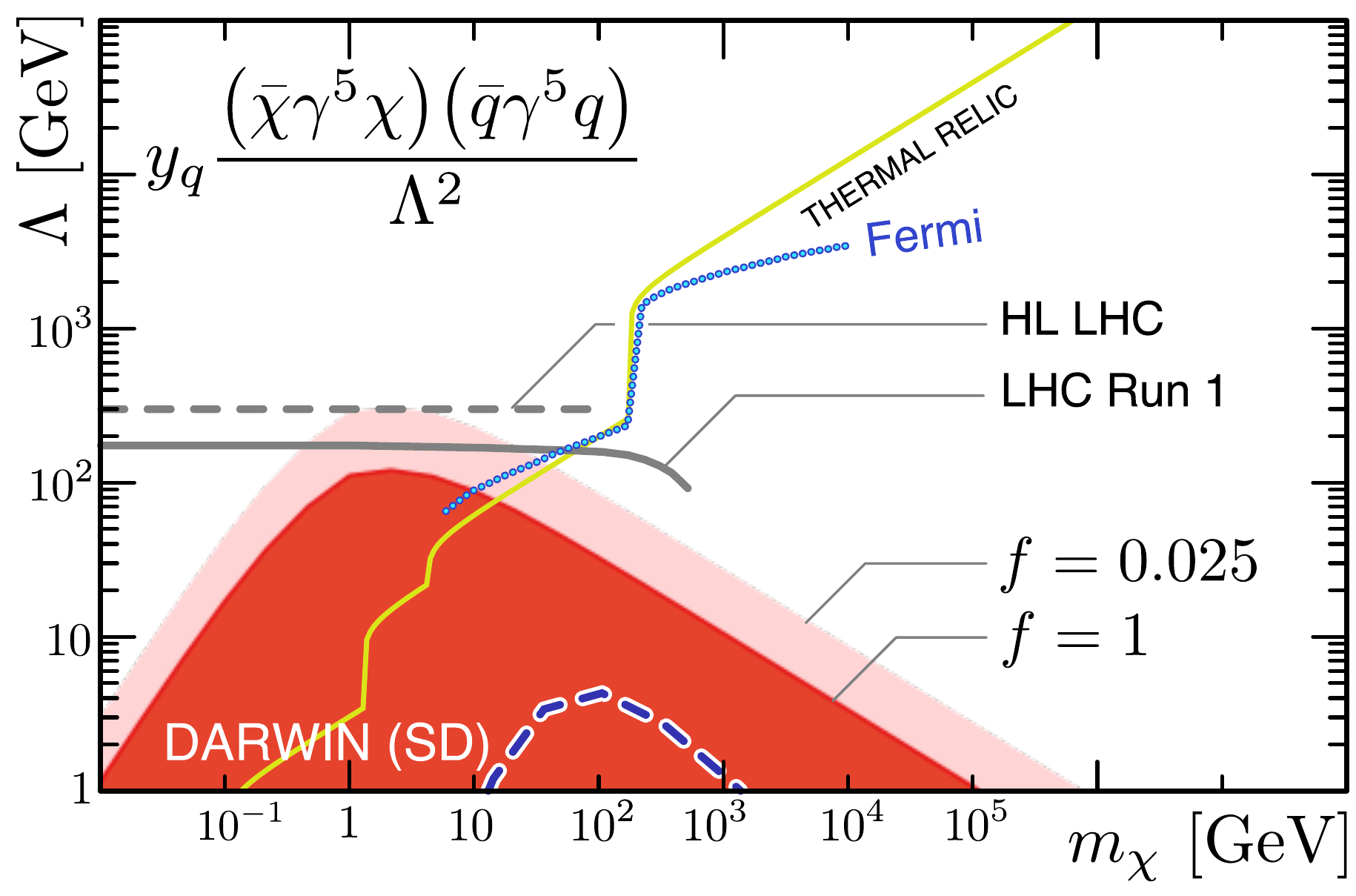} 
\caption{The reach of neutron star heating for operators considered in this work for \DM capture efficiency of $f$ = 1 (solid red) and $f = 0.025$ (pink).
The current and expected reach from direct detection and collider experiments are plotted for comparison.
Direct detection limits on spin-independent (-dependent) scattering are from Xenon1T~\cite{1705.06655} (PandaX~\cite{1611.06553}) in solid blue and the future sensitivities of the \acro{DARWIN} experiment are in dashed blue~\cite{1606.07001}. The current (high-luminosity) \acro{LHC} reach assumes $\sqrt{s}$ = 8~TeV, $\Lc_{\rm int} = 20 \ {\rm fb}^{-1}$ ($\sqrt{s}$ = 14~TeV, $\Lc_{\rm int} = 3000 \ {\rm fb}^{-1}$) and is plotted in solid (dashed) gray~\cite{1503.00691,1407.8257}.
For operators with $s$-wave annihilations we plot bounds from Fermi-\acro{LAT} measurements of dwarf spheroidals with a dotted blue line~\cite{1503.02641}.
}
\label{fig:bounds}
\end{center}
\end{figure*}

Our results are presented in Fig.~\ref{fig:bounds} as red (pink) regions for $f=1$ ($0.025$). 
$f=0.025$ is approximately the smallest capture efficiency  for which kinetic heating is observationally distinguishable from the null hypothesis, $T_{\rm kin} = 700~{\rm K}$, or $T_{\rm ann} = 1000~{\rm K}$ if captured dark matter annihilates.
We compare this observational reach with current and future direct detection and collider searches. The \acro{DARWIN} experiment projects to probe \DM--nucleon cross sections immediately above the neutrino floor~\cite{1606.07001} .
The collider reach is subject to validity of the contact operator treatment; this restricts it to a region $\Lambda \leq \mdm/2\pi$~\cite{1005.3797,1008.1783,1108.1196,1407.8257,1503.00691}. 
One may derive this condition by \acro{UV}-completing our operators with $s$-channel mediators of mass $m_{\rm MED}$ and couplings $g_{\rm SM}$ and $g_{\rm DM}$. The contact operator cutoff is related to the mediator mass according to $m_{\rm MED} = g_{\rm SM} g_{\rm DM} \Lambda$. Coupling perturbativity limits $m_{\rm MED} \leq 4\pi \Lambda$ and the mediator dynamics are negligible so long as $m_{\rm MED} \geq 2 \mdm$.
The yellow line is where the observed observed relic abundance, $\Omega_\chi h^2 = 0.12$, is saturated by thermal freeze-out through \DM annihilations to quarks.
For $\Maa$ and $\Mpp$, where the annihilation is $s$-wave, we also plot in dotted blue upper bounds on $\Lambda$ from the Fermi-\acro{LAT} observations of $\gamma$-rays from dwarf spheroidals assuming only the quark coupling highlighted in each plot.

The qualitative features of Fig.~\ref{fig:bounds} are understood as follows.
For low \DM mass, direct detection is limited by nuclear recoil thresholds and indirect detection is limited by astrophysical backgrounds. For high \DM mass, colliders are limited by their center-of-mass energies and (in-)direct detection is limited by the \DM number density. Neutron star heating, on the other hand is sensitive to a range of \DM masses, and is limited by Pauli blocking at low masses and the requirement of multiple scatters to capture at high masses. The bumps in the thermal relic and Fermi curves are thresholds where new annihilation final states become kinematically accessible.

\section{Discussions and Future Scope}

In this work we examine the reach of neutron star kinetic heating to constrain the cutoff scale of a set of effective contact operators that describe the interactions of Majorana \DM and quarks.
When spin-independent scattering dominates, neutron star heating and underground direct detection give comparable sensitivities, with the former performing better at high \DM masses. 
When spin-dependent scattering dominates, neutron star heating is sensitive to cutoff scales at least an order of magnitude higher.
In the case where spin-dependent scattering is also velocity-suppressed, the difference is even more pronounced because momentum transfers are a factor of 5 larger at neutron stars.
Neutron star heating can probe the elusive pseudoscalar--pseudoscalar operator more stringently than the upcoming \acro{DARWIN} experiment across nine orders of \DM mass.  
The \acro{LHC} complements all these limits at low \DM masses.

Detecting this \DM heating mechanism is a compelling astronomical search~\cite{1704.01577}. 
Sufficiently faint, old, isolated, and nearby neutron stars must first be discovered by their radio pulsing with radio telescopes such as \acro{FAST}~\cite{1105.3794}, following which infrared telescopes such as \acro{JWST}, the Thirty Meter Telescope, or the European Extremely Large Telescope must be pointed at them. 
These telescopes should then observe neutron stars at temperatures 10--100 times lower than the upper limit on the oldest ($t_\text{NS} > 10^8$ years) observed neutron stars \cite{1004.0629}.

There are many opportunities to extend this study.
To begin with, one may generalize to all \DM spins and corresponding interaction structures, and may investigate 
the effect of sub-leading terms in the scattering matrix elements \cite{1707.06998}.
It may also be generalized to include inelastic scattering operators~\cite{1409.0536}; the recoil energies typical in \DM--neutron star scattering can probe GeV mass splittings.
Throughout this study we assumed a direct contact operator interaction with quarks. One may alternatively consider leptophilic models where \DM interacts primarily with leptons at tree-level~\cite{1402.6696,1402.7358}. 
Because roughly a tenth of a neutron star is composed of electrons, one may investigate the role of electron scattering for \DM capture.
While this scenario leads to weak direct detection or collider bounds, the thermal relic and indirect detection bounds remain important.
We leave these investigations for the future and eagerly await first light at Webb.

\section*{Acknowledgements}

We thank Joe Bramante for comments on the manuscript,
and 
Fady Bishara,
Adam Martin
and Tim Tait 
for valuable conversations. 
\acro{NR} is supported by the National Science Foundation under Grant No. \acro{PHY}-1417118.
\acro{HBY} is supported by the \acro{U.S.}~Department of Energy under Grant No.~de-sc0008541 (\acro{HBY}) and the Hellman Fellows Fund.
\acro{NR} and \acro{PT} thank the Perimeter Institute for Theoretical Physics (\acro{PI}) for its support during the ``New Directions in Dark Matter and Neutrino Physics'' workshop, where part of this work was completed.
Research at \acro{PI} is supported by the Government of Canada through the Department of Innovation, Science and Economic Development and by the the Province of Ontario through the Ministry of Research \& Innovation.



\input{references.tex}\end{document}

%% file: universalnewcommands.tex
\newcommand{\lsim}{\lesssim}

\def\Lc{\mathcal{L}}
\def\Oc{\mathcal{O}}

\newcommand{\beq}{\begin{equation}}
\newcommand{\eeq}{\end{equation}}
\newcommand{\bea}{\begin{eqnarray}}
\newcommand{\eea}{\end{eqnarray}}
\newcommand{\nn}{\nonumber}